\documentclass[10pt,letterpaper]{article}
\usepackage[top=0.85in,left=2.75in,footskip=0.75in]{geometry}

\usepackage{changepage}

\usepackage[utf8]{inputenc}
\usepackage{textcomp,marvosym}

\usepackage{fixltx2e}

\usepackage{amsmath,amssymb}

\usepackage{cite}

\usepackage{nameref,hyperref}

\usepackage{microtype}
\DisableLigatures[f]{encoding = *, family = * }
\usepackage[usenames, dvipsnames]{color}
\usepackage{rotating}

\usepackage[aboveskip=1pt,labelfont=bf,labelsep=period,,singlelinecheck=off]{caption}

\makeatletter
\renewcommand{\@biblabel}[1]{\quad#1.}
\makeatother
\date{}
\usepackage{lastpage,fancyhdr,graphicx}
\usepackage{epstopdf}
\fancyheadoffset[L]{2.25in}
\fancyfootoffset[L]{2.25in}
\begin{document}
\vspace*{0.35in}

\begin{flushleft}
{\Large
\textbf\newline{The Kinematics of Plant Nutation Reveals a Simple Relation Between Curvature and the Orientation of Differential Growth}
}
\newline
\\
Renaud Bastien\textsuperscript{1,*},
Yasmine Meroz\textsuperscript{2}
\\
\bigskip
\bf{1} Department of Collective Behaviour, Max Planck Institute for Ornithology and Department of Biology, University of Konstanz, Konstanz, Germany\\
\bf{2} John A. Paulson School of Engineering and Applied Sciences, Harvard University, Cambridge, MA 02138, USA.\\
\bigskip

* rbastien@orn.mpg.de
\end{flushleft}

\section*{Abstract}
Nutation is an oscillatory movement that plants display during their development.  Despite its ubiquity among plants movements, the relation between the observed movement and the underlying biological mechanisms remains unclear. Here we show that the  kinematics of the full organ in 3D gives a simple picture of plant nutation, where the orientation of the curvature along the main axis of the organ aligns with the direction of maximal  differential growth.
Within this framework we reexamine the validity of widely used experimental measurements of the apical tip as markers of growth dynamics. We show that though this relation is correct under certain conditions, it does not generally hold, and is not sufficient to uncover the specific role of each mechanism. As an example we re-interpret previously measured experimental observations using our model.

\section*{Author Summary}
In his writings, Darwin considered nutation, the revolving movement of the apical tip of plants, as the most widespread plant movement \cite{Darwin1880}. In spite of its ubiquity, plant nutation has not received as much attention as other plant movements, and its underlying mechanism remains unclear. A better understanding of this presumably growth-driven process is bound to shed light on basic growth processes in plants. In the work presented here we redefine the problem by describing the kinematics in three dimensions, as opposed to the typical description restricted to the horizontal plane. Within this framework we reveal a simple picture of the underlying dynamics, where the orientation of curvature follows the orientation of maximal differential growth. This parsimonious model recovers the major classes of nutation patterns, as shown both analytically and numerically. We then discuss the limitations of classical measurements where only the movement of the apical tip is tracked, suggesting more adequate measurements.


\section*{Introduction}
During their development, plant organs display a large range of movements. These movements may be broadly divided into two classes; tropisms and nastic movements. Tropisms are the reorientation towards an external stimulus, {\it e.g.} light or gravity \cite{Moulia2009,Christie2013}. Nastic movements account for endogenous, autonomous movements and are not directed towards an external stimulus. Nutation, often called circumnutation, is a particular class of nastic movements, in which the plant organ successively bends in different directions, resulting in an apparent oscillatory swinging motion. Despite its ubiquity among plant movements nutation has not been studied as extensively as tropisms, and the mechanism responsible for this movement, as well as its regulation, remain unclear (see \cite{Brown1993,Migliaccio2013,Mugnai2015} for a review). 

Current theories or concepts of nutational mechanisms generally fall into two categories \cite{Johnsson1997}. 
The first suggests the influence of external drivers such as gravity or light, where the movement stems from an overshoot during the straightening of the plant in response to the direction of gravity (or light). The second assumes an endogenous driver such as an oscillator, suggested by Darwin~\cite{Darwin1880}, possibly related to the growth process itself~\cite{Brown1990,Brown1993}.
Studies have shown that though gravitropism may influence and modify the observed movements, the two processes exist independently~\cite{Brown1993, Israelsson1967,Yoshihara2006}, consistent with symmetry arguments which indicate that gravitropism alone cannot induce movement outside of the plane defined by the main axis of the plant and the direction of gravity~\cite{Bastien2013,Bastien2015}. 
Together with the fact that nutation is observed in the absence of light, this suggests that external cues cannot drive nutation.
Interestingly, Brown~\cite{Brown1993, Mugnai2015} postulated that since nutation does not present any significant evolutionary benefit, it may be the consequence of some fundamental mechanism in the growth process.
 Observations of rice coleoptile mutations ({\it lazy}) that grow normally yet do not exhibit nutation~\cite{Yoshihara2006}, suggest  that   growth  alone may not be sufficient to generate nutation~\cite{Brown1990,Brown1993}.
Therefore based on the present literature,  the strongest hypothesis remains a growth-driven endogenous oscillator.

%
%
  

\begin{figure}[htbp]
  \begin{center}
    \includegraphics[width=0.9\textwidth,]{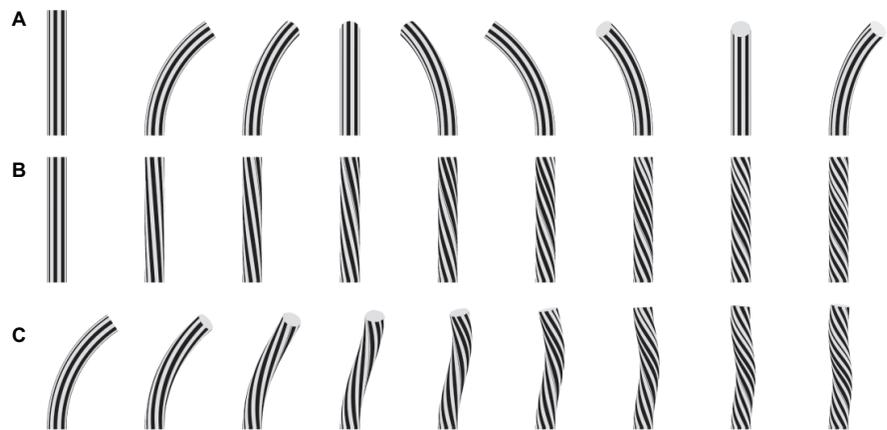}
    \caption{
    {\bf A cylinder is used as a simplified representation of an organ.} Lines parallel to the median line of the organ are drawn on the surface. A. The organ can be curved in different direction of space without material torsion: parallel lines on the surface remain parallel. B. Material torsion of increasing intensity are applied on the organ (from left to right). The torsion does not change the shape of a straight cylinder. However parallel lines at the surface of the organ are no longer parallel to the median line, but are tilted.
C. Material torsion of increasing intensity (from left to right) is applied to a curved organ. The organ does not lie in a plane anymore but the curvature takes different direction in the 3D space. The organ displays a helical shape.
}
    \label{torcurv}
   \end{center}
\end{figure}

We note that curvature of an elongated organ in three dimensional space can result from two different growth mechanisms, namely bending and torsion. Bending can result from the differential growth of the opposite sides of an organ~\cite{Silk1984,Bastien2014}, i.e an initially straight organ will bend towards the direction of minimal growth. Studies have mainly focused on the case where the organ is curved in the same (vertical) plane as that of the differential growth, restricting movement to that plane only. However the plane of curvature should change when it is not in the same plane as that of the differential growth, producing movement in the horizontal (apical) plane. It is instructive to note that lines drawn on the surface of the organ, parallel to its main axis, will remain parallel to the main axis during its growth regardless of the direction of curvature, as shown in  Figure~\ref{torcurv}.A.
The second mechanism, torsion, is responsible for the movement of twining plants \cite{Isnard2009a,Isnard2009b}, and is due to the helical arrangement of cells around the main axis of the plant, possibly due to the torsional arrangement of cellulose \cite{Landrein2013}. In this case parallel lines  drawn on the surface of the organ will take a helical form around the organ during its growth, as shown in Figure~\ref{torcurv}, e.g. the cotyledon on top of a hypocotyl  will rotate. 
However this process can lead to a 3D curved organ only if the organ is already initially curved, and furthermore results in a helical form (see Figure~\ref{torcurv}.C). Moreover, observations of torsion in nutating plant organs have been found to be too slow to account for the observed nutation~\cite{Stolarz2014,Berg1992}. These observations hint that the dominant growth mechanism underlying nutation is differential growth under the action of an internal oscillator~\cite{Berg1992}.
This internal oscillator could then be related to the auxin dynamics or the sensitivity of the membrane to auxin. Indeed a relation has been found between oscillations in ion fluxes and nutation \cite{Shabala1997, Shabala2003}. 
Moreover there are some reports of relationships between nutation and biological rhythms ~\cite{Schuster1997, Buda2003},   demonstrating genetically that the circadian clock controls nutation speed~\cite{Niinuma2005}. Together, these results suggest that genetically regulated rhythmical membrane transport processes are central to plant nutation, and may play the role of an internal oscillator~\cite{Mugnai2015}.

In this study we consider nutation as a growth-driven process,  in line with previous work on tropisms and differential growth \cite{Silk1984,Bastien2013,Bastien2014,Bastien2015} where  mechanical effects such as buckling and instabilities are disregarded. We then focus on the relation between  internal oscillatory growth patterns and the  observed movement. 

Attempts have already been made to develop a mathematical model of nutation, but the  full three-dimensional geometry of the organ has been neglected, resulting in an incomplete kinematic description \cite{Johnsson1973}. The existing models account only for the kinematics of the apical tip, and this has been shown to be insufficient to understand the underlying mechanisms, since geometrical and local effects are neglected \cite{Silk1984,Moulia2009}. 

A similar problem exists in the experimental measurement of nutation, where the full dynamics of  plants in three-dimensional space and  time are rarely taken into account. It is common, instead,  to  track the projection of the apical part of the organ  in the plane orthogonal to the gravitational field  ( defined here as the horizontal plane $P_a$) \cite{Stolarz2014,Stolarz2009}. 
Such measurements carried out on different species and organs \cite{Darwin1880,Stolarz2009} exhibit disorganized patterns (zig-zag shaped), and organized patterns (for instance, elliptical patterns, and the limits of these patterns, {\it e.g.} a circle or a line) \cite{Stolarz2014,Brown1990,kitazawa2008,Schuster1997,Stolarz2008}. 
The interpretation of these measurements remains unclear, since the relation between the differential growth pattern and the kinematics in space and time is not clearly defined. 
  

Here we will state a  growth-driven parsimonious model couched within a  three-dimensional geometrical framework ,   accounting for observed classes of movement patterns. The analysis is done both analytically and through numerical simulations. The model's limitations are also discussed. Lastly, the model is applied to existing experimental observations, and  the relevance of  apical (horizontal) measurements is discussed. The details of all calculations are given in the appendix. In addition, an interactive simulator is available online \cite{nutator}. Predefined solutions  are accessible through the numerical key of the keyboard and are referenced throughout the manuscript.

\section*{Models and Methods}

\subsection*{Geometrical Description}

\begin{figure}[htbp]
  \begin{center}
    \includegraphics[width=0.9\textwidth,]{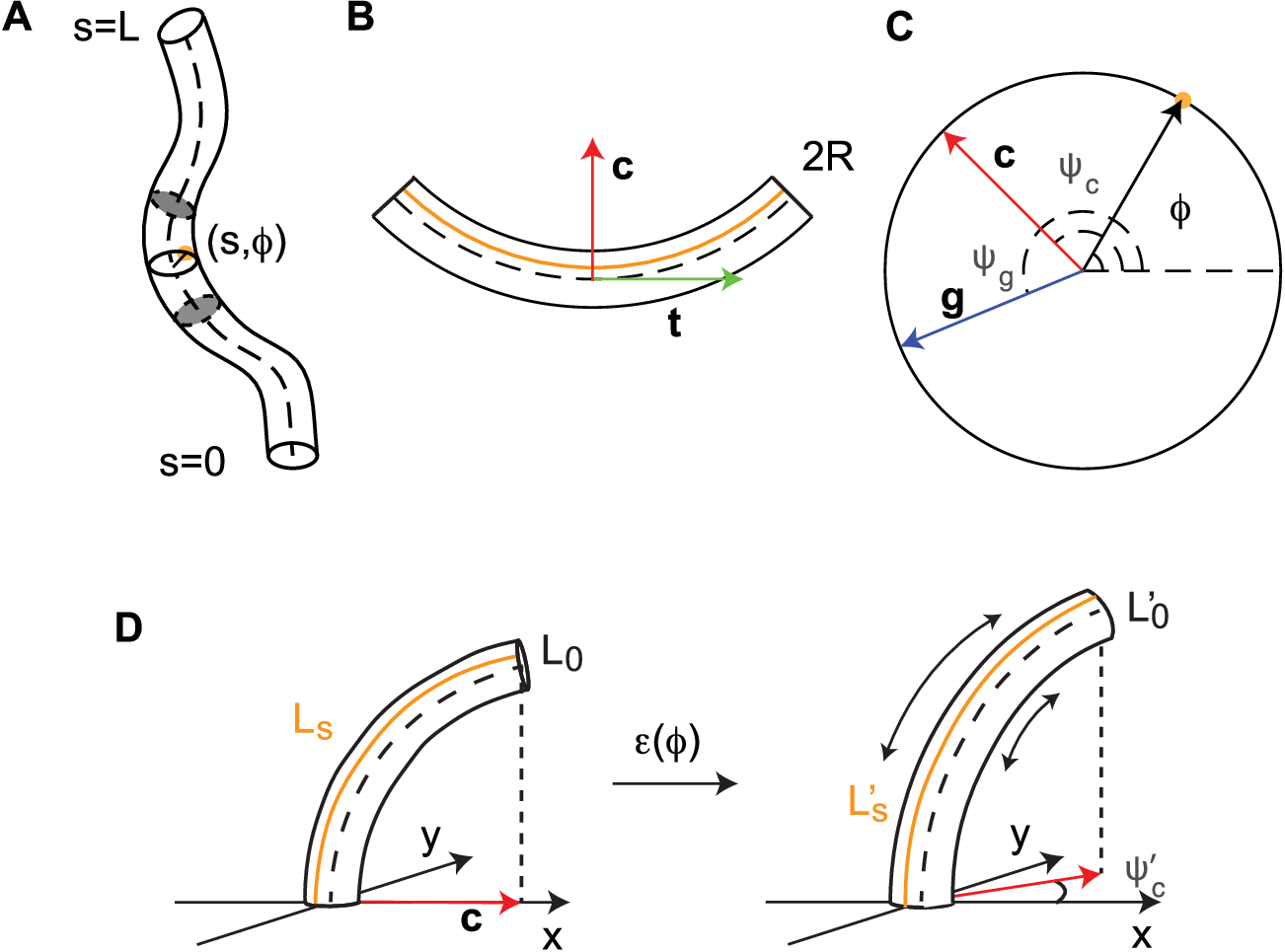}
    \caption{
{\bf A. A plant organ in 3D, described in cylindrical coordinates.} The parameter $s$ runs along the abscissa of the organ, with the base at $s=0$ and apex at $s=L$. A cross-section of of the organ is shown, with a point, in orange, on its circumference defined by the pair $(s, \phi)$. Here $s$ states where the cross-section is along the organ, and $\phi$ determines the orientation from an arbitrarily chosen starting point.
B. An element of the organ shown in A, delimited by two grey cross-sections, projected onto the plane defined by the vectors $\mathbf{t}$ (the tangent, in green), and $\mathbf{c}$ (the normal in the principal direction of curvature, in red). In orange, a segment parallel to the median line is defined by $\phi$ C. The cross section of the organ shown in A in the plane perpendicular to $\mathbf{t}$. Again, an element on the circumference of the organ is defined by the angle $\phi$. The direction of the vector normal in the principal direction of curvature, $\mathbf{c}$ in red, is defined by the angle $\phi=\psi_c$. The principal direction of the differential growth, defined by the vector $\mathbf{g}$ in blue, is defined by the angle $\phi=\psi_g$.Due to the cylindrical symmetry considered for the organ, the direction given by $\phi=0$ is defined arbitrarily but continuously on the whole organ.  When the organ is not curved, the direction given by any $\phi$ defines a straight line along the organ parallel to the main axis of the plant. D. An element cylinder curved in the 3D space, here in the direction $\psi_c=0$. Its median length is given by $L_0$, while a segment on the surface has a length $L_s(\phi)$. After a strain $\epsilon(\phi)$ is applied so that the length of each segment on the surface is now $L'_s(\phi)$. The cylinder is now curved in a different direction, $\psi'_c$, while the curvature is also modified, $C'$. Finally the length of the median line is given by $L'_0$. 
See Movie~S1.}
    \label{torsion}
   \end{center}
\end{figure}
The geometric framework used to describe the kinematics of tropisms~\cite{Bastien2013,Bastien2014,Bastien2015}  is only sufficient to describe growing elongated organs in a single plane and is therefore inadequate here.  Unlike the movements observed for example in gravitropism where the curve is constrained to a unique plane, in nutation the organ is curved along different planes in 3D space (Figure~\ref{torcurv}.A). 

\begin{figure}[htbp]
  \begin{center}
    \includegraphics[width=0.9\textwidth,]{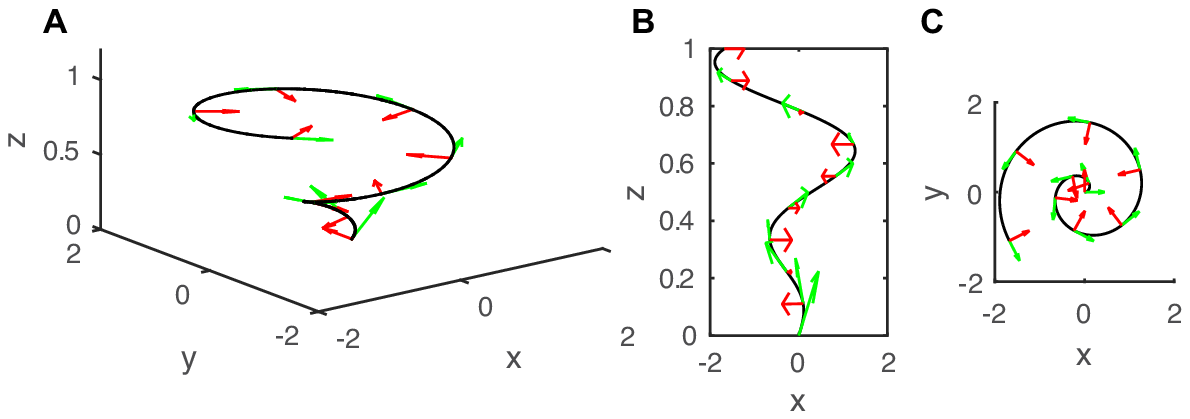}
    \caption{{\bf A. A curve in 3D  space $(x,y,z)$. } Here, a generalized spiral has been chosen as it provides a nice and simple illustration of an organ curved in different planes in 3D. The curve is described at each point by two vectors, as defined in Figure~\ref{torsion}: the tangent and normal to the curve in the plane of the principal direction of curvature, $\mathbf{t}$ and $\mathbf{c}$ (shown in green and red respectively). The vectors are orthogonal to each others. B. and C. present the projections of the curve onto the  $(x,z)$and $(y,z)$ planes respectively. See Movie~S1. }
    \label{vectors}
   \end{center}
\end{figure}

We start by introducing a few assumptions and definitions which will be essential for the  construction of our three dimensional model (Figure~\ref{torcurv}).
 The organ is assumed to be cylindrical with a constant radius $R$ along the organ. It is assumed that no shear growth is observed, so the cross section remains in plane. The organ is described by the curvilinear abscissa $s$ along the median line. Each point at the surface of the organ is then defined via cylindrical coordinates $(s,\phi)$ where $s$ is its position along the abscissa, and $\phi$ is the angle of the point on the cross section,  relative to an arbitrarily chosen direction.  This description is depicted in Figure~\ref{torsion}.A and in movie~S1.
 In order to fully describe the curvature of an organ curved in an arbitrary direction in space, it is first necessary to define  two vectors: $\mathbf{t}$, the tangent to the median line, and $\mathbf{c}$, the normal (perpendicular) to the median line, as shown in Figure~\ref{torsion}.B. The orientation of the latter in the cross section, $\psi_c(t)$, is in the same plane as the principal direction of curvature (see Figure~\ref{vectors}).
This means that for each element of the curve, the curvature is maximal  in the plane defined by the vectors $\mathbf{t}$ and $\mathbf{c}$ (see Figure~\ref{torsion}.B).  
Figure~\ref{torsion}.C shows a cross section of the shoot, by definition in the plane orthogonal to $\mathbf{t}$, defining the orientation $\psi_c$ of the principal direction of curvature $\mathbf{c}$.

\subsection*{Model}

\subsubsection*{Slaving of the principal direction of curvature to direction of maximal growth}

Given the geometrical framework presented here, we now bring a parsimonious model describing nutation kinematics based on basic mechanic and geometric arguments. We recall that for simplicity we assume an organ with constant radius $R$. 
We consider the  elongation strain rate  $\dot{\epsilon}(\phi, s, t)$ defined at each point $(\phi, s)$ along the surface of the organ, describing a  compatible transformation of a cylindrical organ of radius R and curvature $C(s,t)$ in the direction  $\psi_c(s,t)$ into a cylinder of radius $R$ and curvature $C'(s,t)$ in the direction $\psi'_c(s,t)$. 
We now proceed to represent the strain rate  $\dot{\epsilon}(\phi, s, t)$  as a function of the the variation of curvature $\frac{dC(s,t)}{dt}$ and the variation of the principal direction of curvature $\frac{d\psi_c(s,t)}{dt}$ , in turn leading to equations of motion.


We start with an infinitesimal cylindrical element where the curvature $C(s,t) = C $ and its principal direction $\psi_c(s,t)=\psi_c$  are assumed to be 
constant along the median line of length $L_0$ (Figure~\ref{torsion}.D). Since we are discussing an infinitesimal element, we  drop the dependence on $s$, the position along the median line of the whole organ. We also  drop the temporal dependence $t$ for simplicity.
The length of a segment running along the surface of the infinitesimal curved cylinder, parallel to the median line,  depends on its angle $\phi$  relative to the principal direction of curvature $\psi_c$:

\begin{eqnarray}
\label{L1}
L_s(\phi)=L_0\left(1-CR\cos\left(\phi-\psi_c\right)\right).
\end{eqnarray}
We note that for a line in the principal direction of curvature, i.e. at the inner part of the curve,  the length is minimal, $L_s(\phi=\psi_c) = L_0(1-CR)$, while it is maximal for a line in  the outer part of the curve $L_s(\phi=\psi_c + \pi) = L_0(1+CR)$. A similar relation holds for the cylinder after deformation, namely $L'_s(\phi)=L'_0\left(1-C'R\cos\left(\phi-\psi_c'\right)\right)$.

We note that the elongation strain $\epsilon(\phi)$ at a specific point along the cylinder is defined as the ratio between the elongation length $L'_s(\phi)-L_s(\phi)$ and the original length $L_s(\phi)$
\begin{equation}
\label{ELL}
\epsilon(\phi) = L'_s(\phi)/L_s(\phi) - 1
\end{equation}
Similarly the  average elongation strain rate defined as 
\begin{equation}\label{E}
E = \frac{1}{2\pi}\int_{-\pi}^{\pi}{d\phi \epsilon(\phi)}
\end{equation}
is related to the ratio of the median lengths. 
Substituting $L_0' =  L_0(E+1)$ , the relation for the cylinder after deformation reads:
\begin{eqnarray}
\label{L2}
L'_s(\phi)=(1+E)L_0\left(1-C'R\cos\left(\phi-\psi_c'\right)\right).
\end{eqnarray}
Substituting   equations~\ref{L1} and~\ref{L2} into equation~\ref{ELL} yields  an expression for the elongation strain $\epsilon(\phi)$ :
\begin{eqnarray}\label{epsilon_phi_E}
\epsilon(\phi)=(1+E)\frac{1-C'R\cos(\phi-\psi_c')}{1-CR\cos(\phi-\psi_c)}-1.
\end{eqnarray}
We now introduce time by considering an infinitesimal time step $dt$ and substituting the  first order differentials $\epsilon(\phi) = \dot{\epsilon}(\phi) dt$, $E(\phi) = \dot{E}(\phi) dt$, $C' = C+dC$, and $\psi'_c = \psi_c + d\psi_c$,  assuming that terms with second order infinitesimals are negligible, and noting that $\cos{(\phi-(\psi_c+d\psi_c) )} = \cos{(\phi-\psi_c)} + \sin{(\phi-\psi_c)d\psi_c }$ . We use the common dot notation for strain rates. Rearranging and reintroducing the explicit dependence on $t$  into the notation, equation~\ref{epsilon_phi_E} now reads:

\begin{equation}
\label{EE4}
\dot{\epsilon}(\phi, t) = \dot{E}(t) + \frac{\textstyle \sin\left(\phi-\psi_c(t)\right)\frac{\textstyle d\psi_c(t)}{\textstyle  dt} C(t)R 
+ R\cos\left(\phi-\psi_c(t)\right)\frac{\textstyle dC(t)R}{\textstyle dt}}{1-C(t)R\cos\left(\phi-\psi_c(t)\right)}.
\end{equation}
This equation relates the elongation rate of a fiber, or a segment running along the organ surface, to the rate of change in curvature magnitude $\frac{\textstyle dC(t)}{\textstyle dt}$ and direction $\frac{\textstyle d\psi_c}{\textstyle dt}$. 

We recall that this relation does not exhibit an explicit dependence on $s$ since the argument was for an infinitesimal cylindrical element. The full organ is then made up of consecutive deforming (elongating) infinitesimal elements, each with dynamics described by equation~\ref{EE4}. We note that one cannot just yet reintroduce the dependence on $s$, the position along the median of the organ relative to the base, since as the organ deforms and elongates the positions of the constituting elements move along the organ. The variable $s$ is relative to the base only, and not  to the material elements. Therefore when considering the whole organ we introduce the material derivative, co-moving with each element of the organ~\cite{Silk1984,Moulia2009,Bastien2014} :
\begin{equation}
\label{DT}
\frac{D}{Dt}=\frac{\partial}{\partial t}+v(s,t)\frac{\partial}{\partial s }
\end{equation}
where $v(s,t)$ is the velocity of the average growth-induced displacement of each element at $s$ at time $t$, and is defined as the integral of the average elongation rate on the median line $\dot{E}(s,t)$ defined in equation~\ref{E}:
\begin{equation}
v(s,t)=\int_0^sds^\prime \dot{E}(s^\prime,t).
\end{equation}
Equation~\ref{EE4} can  be modified to account for a cylindrical organ of constant radius, that is elongating and where the curvature at each point along the median line of the organ $s$ is modified in intensity and direction.
We now rewrite equation~\ref{EE4} to account for the elongation strain rate along the whole organ by replacing  time derivatives with the material derivative in equation~\ref{DT}, and reintroducing the explicit dependence on $s$: 
\begin{equation}
\label{EE3}
\dot{\epsilon}(\phi,s,t) = \dot{E}(s,t) + \frac{\textstyle \sin\left(\phi-\psi_c(s,t)\right)\frac{\textstyle  D \psi_c(s,t)}{\textstyle Dt} C(s,t)R + \cos\left(\phi-\psi_c(s,t)\right)\frac{\textstyle D C(s,t)R}{\textstyle Dt} }{1-C(s,t)R\cos\left(\phi-\psi_c(s,t)\right)}.
\end{equation}
Let us note that equation \ref{EE3} has two main contributions that dominate in orthogonal planes. In the plane parallel to the principal direction of curvature, i.e $\phi = \psi_c$ or $\phi = \psi_c + \pi$, we have $\sin(\phi-\psi_c(s,t))=0$ and $\cos(\phi-\psi_c(s,t))=\pm 1$ , therefore the elongation strain rate is only affected by the change in curvature $\frac{\textstyle DC(s,t)}{\textstyle Dt}$ . In the orthogonal plane, i.e $\phi = \psi_c \pm \pi/2$, the elongation strain rate is only affected by the change in the orientation of the curvature plane $\frac{\textstyle D\psi_c}{\textstyle Dt}$.
It is instructive to consider this in terms of differential growth, generally defined as the difference  between elongation strain rates at opposite sides of the organ, divided by the average strain rate:
\begin{equation}\label{eq:diff_growth}
\Delta(\phi,s,t) \equiv \frac{\dot{\epsilon}(\phi,s,t)-\dot{\epsilon}(\phi+\pi,s,t)}{2\dot{E}(s,t)}
\end{equation}
Note that  $\Delta(\phi,s,t)=0$ implies that growth is the same on either side in this plane, while $\Delta(\phi,s,t)=\pm1$ means that all of the growth occurs on one side of the organ or the other.
We can now define the differential growth in the plane parallel to the principal direction of curvature $\psi_c$:
\begin{equation}
\Delta_{\parallel}(s,t) \equiv \Delta(\psi_c,s,t)=\frac{\dot{\epsilon}(\psi_c(s,t),s,t)-\dot{\epsilon}(\psi_c(s,t)+\pi,s,t)}{\dot{E}(s,t)}, 
\end{equation}
and in the orthogonal plane:
\begin{equation} 
\Delta_\perp(s,t) \equiv \Delta(\psi_c+\pi/2,s,t)=\frac{\dot{\epsilon}(\psi_c(s,t)+\pi/2, s, t)-\dot{\epsilon}(\psi_c(s,t)+3\pi/2, s, t)}{\dot{E}(s,t)}.
\end{equation}

Substituting equation \ref{EE3}  and rewriting,  leads to an equivalent set of equations:
\begin{eqnarray}
\frac{DC(s,t)R}{Dt}&=&\Delta_\parallel(s,t)\dot{E}(s,t)(1-C(s,t)^2R^2)\sim\Delta_\parallel(s,t)\dot{E}(s,t)\label{dcpa3}\\
 \frac{D\psi_c(s,t)}{Dt} &=&\Delta_\perp(s,t)\frac{\dot{E}(s,t)}{C(s,t)R}\label{dcpe3}
\end{eqnarray}
Assuming the radius of curvature $1/C(s, t)$ remains large compared to the radius $R$ of the organ, {\it i.e.} $C(s,t)R \ll 1$, the quadratic prefactor in equation \ref{dcpa3}, $(1-C(s,t)^2R^2)$, can be neglected \cite{Bastien2014}.

Equation~\ref{dcpa3} expresses the variation of curvature as a function of $\Delta_\parallel$ the differential growth in the plane of the principal direction of curvature only, and is identical to the equation found in the case of in-plane curvature \cite{Bastien2014,Silk1984}. Out of plane curvature is then fully described by the addition of equation~\ref{dcpe3}, which  expresses the variation of the principal direction of curvature as a function of $\Delta_\perp$ the differential growth in the orthogonal plane. Moreover we see that only three quantities govern the relation between growth and the resulting nutation movement: the average elongation rate on the median line $\dot{E}(s,t)$, the differential growth in the plane of curvature $\Delta_{\parallel}(s,t)$, which expresses the curvature variation {\it via} equation \ref{dcpa3},  and finally the differential growth in the orthogonal plane $\Delta_\perp(s,t)$, which expresses the change of orientation of the plane of curvature {\it via} Equation \ref{dcpe3}.

It is instructive to note that, without loss of generality, it is possible to express $\Delta_\parallel(s,t)$ and $\Delta_\perp(s,t)$  as a projections  of the principal direction of differential growth $\Delta\left(\psi_g(s,t)\right)$ , {\it i.e.} where the differential growth is  maximal:
\begin{equation}
\label{dpeg}
\Delta_{\parallel}(s,t)=\Delta\left(\psi_g(s,t)\right)\cos(\psi_g(s,t)-\psi_c(s,t)),
\end{equation}
\begin{equation}
\label{dpag} 
\Delta_\perp(s,t)=\Delta\left(\psi_g(s,t)\right)\sin(\psi_g(s,t)-\psi_c(s,t)).
\end{equation}

In summary, the model presented here suggests that nutation results from the slaving of the direction of maximal curvature, $\psi_c(s, t)$, to the direction of maximal growth $\psi_g(s,t)$.

\subsubsection*{The effects of growth}

In linear organs growth is often limited to a zone below the apex \cite{Bastien2014,Silk1984}. Two cases are usually considered: i. when the length $L$ of the organ is smaller than the length of the growth zone $L<L_{gz}$, the organ elongates along its entire length, ii. when $L>L_{gz}$ the growth is localized to a subapical zone.

The effects of growth on the variation of curvature,  restricted to the plane of curvature, equation~\ref{dcpa3}, have been discussed in \cite{Bastien2014}. Two main destabilizing effects of the growth process have been described: i. A passive orientation drift, where the organ elongates in absence of differential growth, $\dot{E}(s,t)>0$ and $\Delta_{\parallel}(s,t)=0$, so the curvature of the organ remains the same while the length of the organ is modified. The angle of the organ, relative to the vertical, passively drifts within the plane of curvature during growth. ii. A fixed curvature; 
when an element close to the base leaves the growth zone,  $\dot{E}(s,t)=0$, the curvature cannot be modified there anymore.
It has been shown experimentally that these effects are negligible due to regulation governed by proprioception~\cite{Bastien2014}.

Similar effects are expected to take place when the variation of the direction of curvature is considered, as in equation~\ref{dcpe3}. i. In the absence of regulation, $\Delta(\phi)=0$, the direction of curvature cannot be modified anymore. If the organ is curved in a single plane, no modification will be observed of the shape or movement. However if an organ is curved in multiple planes, displaying a helical shape,

The radius of the helix will increase. ii.  As elements leave the growth zone, the direction of curvature will remain fixed. If the direction of curvature is oscillating, this will result in a helical shape of the fixed part of the plants.
However those destabilizing effect does not hold the same role in the postural control of the plant, the ability of the organ to reach the vertical position and align with the direction of gravity~\cite{Moulia2006,Bastien2013}. 

%
%
As {\it a priori} there is no preferred direction of curvature, 
only the magnitude of curvature  should be regulated, and not its direction. This regulation depends mainly on proprioception.

The effects of elongation can be studied as a perturbation~\cite{Bastien2014}, and it is therefore proposed first to neglect the elongation of the stem on the resulting pattern, so that $\dot{E}$ and the length of the organ $L=L_0$ are constant. Once the movement is clearly defined without growth, it will be simple to discuss the perturbation due to growth.

\subsubsection*{Measurements of the apical tip in the horizontal plane}

As mentioned earlier, it is common to measure nutation by tracking the projection of the apical part of the organ in the horizontal plane \cite{Stolarz2014,Stolarz2009}. In order to understand these measurements  we first clearly state a sufficient and parsimonious set of hypotheses relating whole organ dynamics to the apical movement in the horizontal plane (which are otherwise generally implicitly assumed). We propose the following:

 H1 - The dynamics are homogeneous along the organ: the dynamics of the differential growth do not depend on the local position along the organ and are then considered to be the same over the whole organ $\Delta(\psi,s)=\Delta(\psi)$.

 H2 - There is a unique relation between the shape of the organ and the position of the apical tip.  As many 3D shapes  can result in the same position of the apical tip in the horizontal plane, constraints on the shape need to be stated explicitly so that only one 3D shape of an organ can be mapped to a single position of the apical tip in the horizontal plane.

H3 - If the projection is only measured on the horizontal plane, the effects due to the elongation of the organ may be disregarded. 

With these three hypotheses the entire organ is considered as a whole. The associated curvature and differential  growth  considered are give by their respective combined effects on all parts of the organ. 
%
%

\subsubsection*{Variation of the principal direction of the differential growth $\psi_g(s,t)$ as an internal oscillator}


The variation of the principal direction of the differential growth $\psi_g(s,t)$ provides a natural way to implement an endogenous oscillator, since it is the only known growth driven process occurring parallel to the horizontal plane.
To date there are no exact experimental observations of the temporal variation of  $\psi_g(t)$, however the existence of a linear oscillator seems consistent with  measurements made on opposite sides of an organ \cite{Berg1992,Baskin2007}, where the differential  growth  oscillates periodically from one side to the other. 

Moreover, the model allows to extract $\psi_g(s,t)$ and $\Delta(\psi_g(t),t)\dot{E}(t)(s,t)$ from 3D experimental data of the curvature $C(s,t)$ (and therefore also $R$ and $\psi_c(s,t)$).  Substituting equations~\ref{dcpa3} and~\ref{dcpe3} in equations~\ref{dpeg} and~\ref{dpag} leads to the following relations: 
%
%
\begin{equation}
\psi_g(s,t)-\psi_c(s,t)=\arctan\left(\frac{\textstyle C(s,t)R\frac{\textstyle D\psi_c(s,t)}{\textstyle Dt} }{\frac{\textstyle DC(s,t)R}{\textstyle Dt}}\right)\label{psig}
\end{equation}
and
\begin{equation}
\Delta\left(\psi_g(s,t),s,t\right)\dot{E}(s,t)=\sqrt{\left(\frac{\textstyle DC(s,t)R}{\textstyle Dt}\right)^2+\left(C(s,t)R \frac{\textstyle D\psi_c(s,t)}{\textstyle Dt}\right)^2}.\label{DE}
\end{equation}
Considering Equations~\ref{dcpa3} and~\ref{dcpe3} we note that for an oscillatory movement such as nutation, the direction of curvature changes over time, i.e. $\frac{D\psi_c(s,t)}{Dt} \neq 0$. In absence of compression the median elongation rate $\dot{E}(s,t)$ is always positive, and 
cannot  control the direction of $\psi_c(s,t)$ \cite{Bastien2014}. Therefore the oscillatory movement can only be governed by $\Delta_\perp(s,t)$ and $\Delta_\parallel(s,t)$.


We now consider two basic cases for the functional form of $\psi_g(s,t)$, and analyze the ensuing organ movement.
The simplest case is given when the direction of the differential growth is  fixed, $\psi_g(s,t)=\psi_g$. The orientation of the organ is modified so that the principal direction of curvature $\psi_c(s,t)$ aligns with the direction of the differential growth $\psi_c(s,t)$ (see Appendix section~1.1). Despite its triviality, this result sheds light on the behavior of the out-of-plane curvature driven by differential growth. The organ tries to align, following the main direction of the differential growth, {\it i.e.} $\psi_c(s,t)\rightarrow\psi_g(s,t)$. As can be seen from equations~\ref{dcpa3} and~\ref{dcpe3}, once an organ is aligned, i.e. $\psi_c(s,t) = \psi_g(s,t)$, there is no movement outside of this plane and only vertical bending is observed. This is also confirmed numerically (shown in Movie~S2), where simulations use a an initial curvature and  principal direction constant along the organ, $C(s,0)=C_0$ and $\psi_c(s,0)=\psi_0$ (see \cite{nutator} - key 2). The details of the simulations can be found in the caption of Movie~S1.

We now consider a more complex case, where the orientation of the differential growth  rotates periodically with a constant angular frequency $\omega$. The direction of the differential growth is then given by
\begin{equation}
\psi_g(s,t)=\omega t.
\end{equation}
In this case an analytical stability analysis can be performed. The movement of a single element displays a periodic movement, the  periodicity of which is given by the direction of the differential growth $\psi_g(t)$ (see Appendix section~1.2 and Figure~S1 therein). Furthermore the stability and periodicity are independent of the initial conditions. This means that even when the rotation is  not centered around the base of the organ, the pattern remains stable and the periodicity is still given by the internal oscillator (simulations giving rise to a circular pattern are shown in Movie~S3).

\begin{figure}[htbp]
  \begin{center}
    \includegraphics[width=0.9\textwidth,]{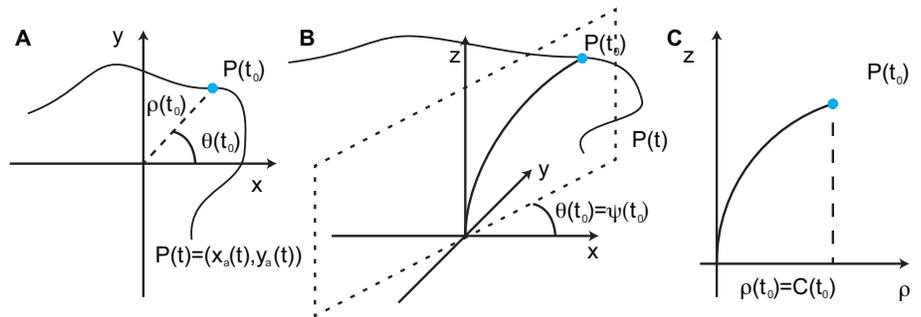}
    \caption{{\bf A.  The trajectory of the  apical tip projected onto the $(x,y)$ plane produces a parametric curve $P(t)=(x_a(t),y_a(t))$. } Under the assumptions H1-H3, the position of the apical projection $P(t)$, gives information on the amplitude and  the direction of the plane of curvature, in polar coordinates $(\rho(t),\theta(t_0))$. B. At a given time $t_0$ the organ is considered to be curved with a constant curvature $C(t_0)$ along the organ in a single plane defined by the angle $\psi(t_0)=\theta(t_0)$. The measurement of $\theta(t)$, the angle of the projected apical curve,  is then equivalent to $\psi_c(t)$ C. Inside the plane perpendicular to the horizontal plane that contains the organ, the curvature can be measured directly as the value of $\rho(t_0)$ (see Appendix section~2-4 for a mathematical justification and a description of the approximations).  }
    \label{geo2}
   \end{center}
\end{figure}

As mentioned earlier, existing experimental observations concerning nutation measure the movement of the apical tip in the horizontal  $(x,y)$  plane, in the form of a parametric curve $P(t)=(x_a(t),y_a(t))$. In order to interpret existing data, we analyze the projected movement in the context of our suggested model.
Under the hypotheses H1-H3 measurements in the horizontal plane should give  direct information on the dynamics of the plant if its shape is known. According to H1, the simplest case is considered, where the dynamics do not depend on the local position along the organ,  the curvature is the same along the organ, $C(s,t)=C(t)$ and the entire organ is curved inside the same plane, $\psi_c(s,t)=\psi_c(t)$ . Therefore there is no dependence on space, and the material derivative $\frac{ D}{ Dt}$, in equation~\ref{DT}, is equivalent to the partial derivative $\frac{ \partial}{ \partial t}$. The projection of the apical tip in the horizontal plane  $(x_a(t), y_a(t))$, or $(\rho(t),\theta(t))$  in polar coordinates  (Figure~\ref{geo2}), is then a direct approximation of $\psi_c(t)$ and $C(t)$ since by definition $\psi_c(t)=\theta(t)$ (Figure~\ref{geo2}) and $C(t)=\rho(t)$ (see Appendix~3). A direct estimate of the principal direction of the differential growth $\psi_g(t)$ can be obtained from the coordinates $(x_a(t),y_a(t))$ of the projection of the apical tip in the horizontal plane (see Appendix~4):

\begin{equation}
\label{psig45}
\psi_g(t)=\arctan\left(\frac{d x_a(t)}{d y_a(t)}\right),
\end{equation}
as well as an estimate of the differential growth term (see Appendix~4):
%
%
\begin{equation}\label{dpsig_de}
\Delta(\psi_g(t),t)\Dot{E}(t)=\frac{2R}{L}\sqrt{\left(\frac{d x_a(t)}{d t}\right)^2+\left(\frac{d y_a(t)}{d t}\right)^2},
\end{equation}

where $L$ is the length of the organ. In most of the published data, $L$ is not available, meaning that values of $\Delta(\psi_g(t),t)\dot{E}(t)$ measured from the horizontal plane can only be compared qualitatively up to a prefactor. On the other hand Equation~\ref{psig45} is independent of $L$, and the principal direction of growth $\psi_g(t)$ can be measured quantitatively from the observed pattern.

\subsection*{Results}

\begin{figure}[htbp]
  \begin{center}
    \includegraphics[width=0.9\textwidth,]{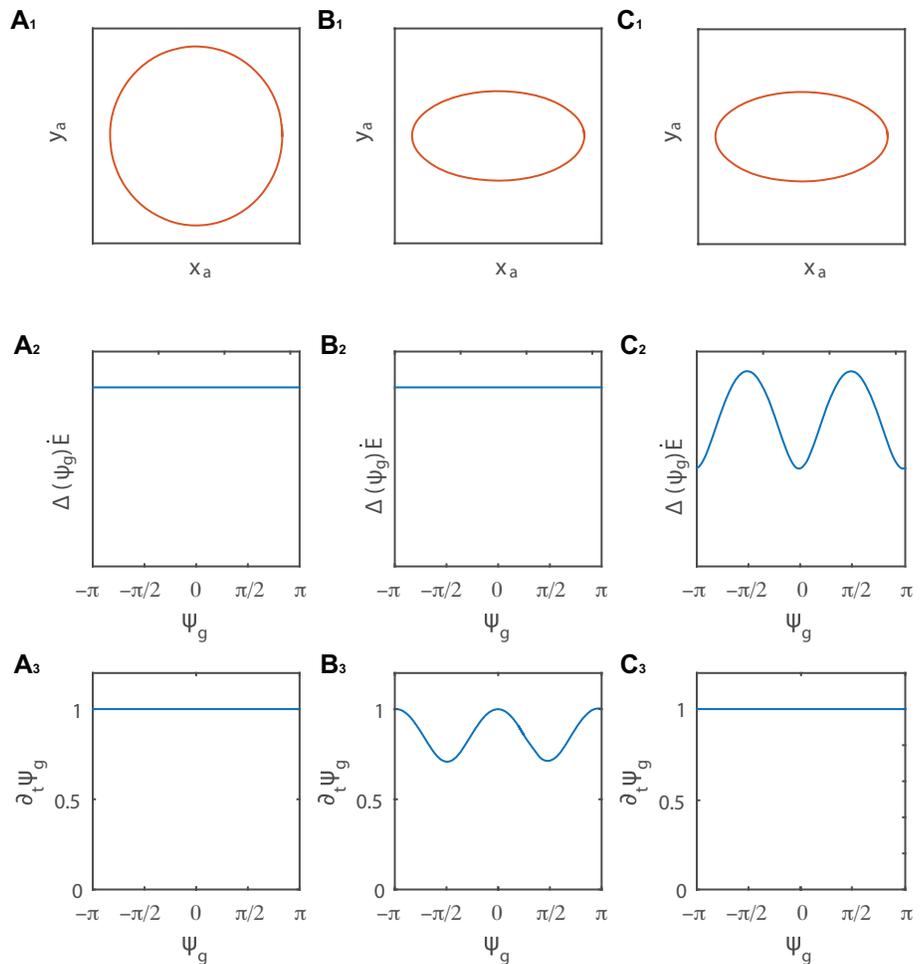}
    \caption{{\bf A. Mechanism underlying  a circular pattern: $A_1$.} A circular pattern in the plane of the apical tip defined by $(x_a(t), y_a(t))$.  This is realized by a  constant differential growth $\Delta(\psi_g(t),t)\dot{E}(t)$ ($A_2$) and constant variation of the direction of the differential growth $\frac{d\psi_g(t)}{d t}$ ($A_3$) . A simulation of an organ with these conditions can be found in Movie~S3. B and C (see \cite{nutator} - key 0 and 1). Underlying mechanisms for an elliptic pattern: $B_1$ and $C_1$. An ellipse in the horizontal plane. This pattern can be realized with a constant constant $\Delta(\psi_g(t),t)\dot{E}(t)$ ($B_2$) and a periodically varying  $\frac{d\psi_g(t)}{d t} $ which is maximal close to the major axis of the ellipse ($B_3$). A simulation can be found in Movie~S4. Another option is if $\Delta\dot{E}$ varies periodically and is maximal close to the major axis of the ellipse ($C_2$), even if  $\frac{d\psi_g(t)}{d t}$ is constant in time ($C_3$). A simulation is presented in Movie~S5.  A combination of these two limiting mechanisms will also lead to an ellipse.}
    \label{shape1}
   \end{center}
\end{figure}

In what follows we re-examine existing experimental observations in the context of our model. We analyze different classes of movements recorded in the horizontal plane, examining possible underlying mechanisms. We first consider the simulated apical trajectories of the most common observed patterns, the circle and the ellipse. 
Figure~\ref{shape1} presents the underlying form of the variation of the principal direction of growth $\frac{d\psi_g(t)}{d t}$ and its differential growth $\Delta(\psi_g(t),t)\dot{E}(t)$, as imposed by the model {\it via} equations~\ref{psig45} and~\ref{dpsig_de}. In the case when the apical tip draws a circle in the horizontal plane, it follows that  $\frac{d\psi_g(t)}{d t}$ and $\Delta(\psi_g(t),t)\dot{E}(t)$ are constant in time (shown in Figure~\ref{shape1}A). In the case of an ellipse, there are two possible mechanisms:  (i) a periodic  $\frac{d\psi_g(t)}{d t}$ with maxima at some $\psi_g^0$ and $\psi_g^0+\pi$
(Figure~\ref{shape1}B), meaning that the direction of differential growth changes faster on opposite sides of the organ, giving less time for the organ to curve out in those directions, resulting in a smaller radius at those ends.  (ii) 
a periodic $\Delta(\psi_g(t),t)\dot{E}(t)$ (Figure~\ref{shape1}C) with maxima at some $\psi_g^0+\pi/2$ and $\psi_g^0+\pi/2$, meaning that the differential growth is larger at opposite sides of the organ, resulting in a greater curvature and therefore also larger radius at those ends (simulations giving rise to these patterns are given in Movies~S3,~S4 and~S5).

For the sake of intuition, Figure~\ref{shape1} presents limit cases where either $\frac{d\psi_g(t)}{d t}$ or $\Delta(\psi_g(t),t)\dot{E}(t)$ are periodic while the other is constant, however in reality both may be periodic, and their relative magnitude and shift in phase will dictate the final apical pattern.
We also note that taking the average value of $\frac{d\psi_g(t)}{d t}$ yields the time it takes for the direction of differential growth to make a full rotation of the organ, i.e. 
\begin{equation}\label{tr}
T_r = 2\pi/\langle \frac{d\psi_g(t)}{d t} \rangle .
\end{equation}
If $\frac{d\psi_g(t)}{d t}$ itself exhibits periodicity, we expect the time between two maxima to coincide with this period, since consecutive maxima are expected to be an angle of $\pi$ apart.
Moreover, we note that $\frac{d\psi_g(t)}{d t}$ and $\Delta(\psi_g(t),t)\dot{E}(t)$ are plotted here as a function of $\psi_g$, assuming the organ does not exhibit torsion, 
i.e. the rotation of the cotyledon on top of the organ's movement,  in which case the behavior would be shifted leading to erroneous conclusions. On the other hand a periodic or constant behavior would still be observed when plotting these values as a function of time. 
Lastly, as the sign of $\Delta(\psi_g(t),t)\dot{E}(t)$ does not change, one cannot distinguish between variation of the median elongation, which is supposed to be positive, and variation of the differential growth term.

\begin{figure}[htbp]
  \begin{center}
    \includegraphics[width=0.7\textwidth,]{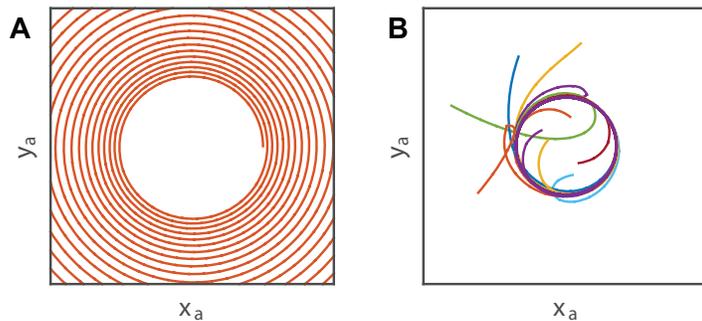}
    \caption{ {\bf $A.$ The effect of growth on the apical movement in the horizontal plane.} A spiral is observed due to the increase in size of the organ. $B$ When proprioception is added to a circular pattern, (see Figure~\ref{shape1}.A), the apical tip modifies its trajectory to turn on a circular pattern centered on the base of the organ. Each colors represent different initial conditions.}
    \label{growth}
   \end{center}
\end{figure}

Let us now consider the effect of elongation on the observed patterns.
In the case where the whole organ is growing,  $L<L_{gz}$, the curvature does not change as the organ grows, but the increasing length of the organ results in a spiral (Figure~\ref{growth}.A). The time to make a full turn remains unchanged, as equation~\ref{tr} is independent of the length of the organ (see \cite{nutator} - key 4). In the case where $L>L_{gz}$, the pattern remains circular, however  the part of the organ outside of the growth zone is fixed in a helical pattern, and the organ is curved in different directions. The observed circular pattern will exhibit a drift. If this helix is small, $C L_{gz}\ll R/L_{gz}$, it may not be noticeable experimentally (see \cite{nutator} - key 5).

No experimental account of this kind of helical pattern has been reported, suggesting a strong regulation of the curvature. Following the case of gravitropism~\cite{Bastien2014}, proprioception is a good candidate for curvature regulation. A proprioceptive term can easily be added to equation~\ref{dcpa3}:

\begin{equation}
\Delta_\parallel(s,t)=-\gamma C+\Delta\left(\psi_g(s,t)\right)\sin(\psi_g(s,t)-\psi_c(s,t))
\end{equation}

The results obtained for a circular pattern are then slightly modified (Figure~\ref{growth}.B). 
Here the curvature of the organ is reduced by proprioception, which tends to straighten the organ \cite{Bastien2013}. The observed pattern is shifted around the base of the organ, in order to reduce the maximal curvature reached by the organ. The apical tip now converges towards a single stable orbit centered around the base that is fully independent of the initial conditions (see \cite{nutator} - key 3).
Due to the evidence that proprioception prevents fixed curvature in the case of gravitropism~\cite{Bastien2014}, it is reasonable to postulate that such regulation is also sufficient in the case of nutation.

\begin{figure}[htbp]
  \begin{center}
    \includegraphics[width=0.7\textwidth,]{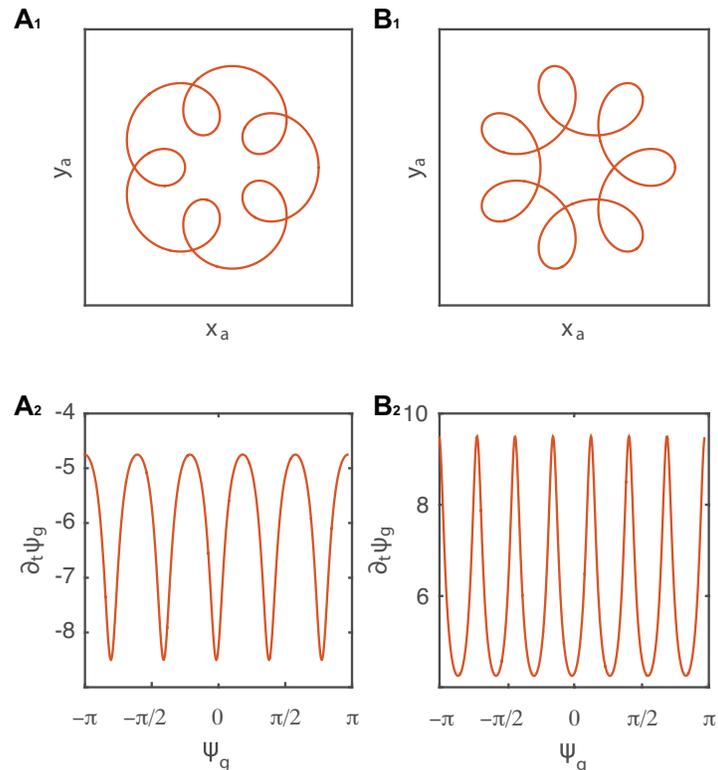}
    \caption{ {\bf Epi- and hypo-trochoids  constructed as a sum of two linear oscillators with angular frequencies respectively $\omega_1$ and $\omega_2$.} $A_1$ $\omega_2=6\omega_1$ and $B_1$ $\omega_2=-6\omega_1$. Plotting $\partial_t\psi_g$ as a function of time is governed by the oscillator with the higher frequency, in this case $\omega_2$ ($A_2$ and $B_2$). Simulations of these patterns are presented in Movies~S5,~S6 and~S7. The analysis of $\frac{d\psi_g(t)}{d t}$ using equation~\ref{psig45} does not allow to extract accurately the values of $\omega_1$ and $\omega_2$, but rather gives and {\it effective} oscillator. }
    \label{spiro}
   \end{center}
\end{figure}

Some experimental observations have shown the existence of epi- and hypo-trochoid patterns (spirograph pattern) \cite{Schuster1997}. These patterns provide an interesting case where the validity of hypothesis H1 (which assumes no local effects) is put in question. Mathematically, a trochoid is constructed as a sum of linear oscillators. If two segments of the organ of length $L_1$ and $L_2$ possess different temporal behaviors of the orientation of differential growth $\psi_{g1}(t)=\omega_1 t$ and $\psi_{g2} (t) =\omega_2 t$, a trochoid will then be observed in the horizontal plane, as shown in Figure~\ref{spiro}, and from the simulations presented in Movies S6 and S7.
Applying equation~\ref{psig45} to the apical curves cannot discern between the two separate oscillators, and will therefore result in an {\it effective} $\psi_g(t)$.
Furthermore, the sign of the effective $\frac{d\psi_g(t)}{d t}$  is dominated by the faster oscillator.
 
\begin{figure}[htbp]
  \begin{center}
    \includegraphics[width=0.9\textwidth,]{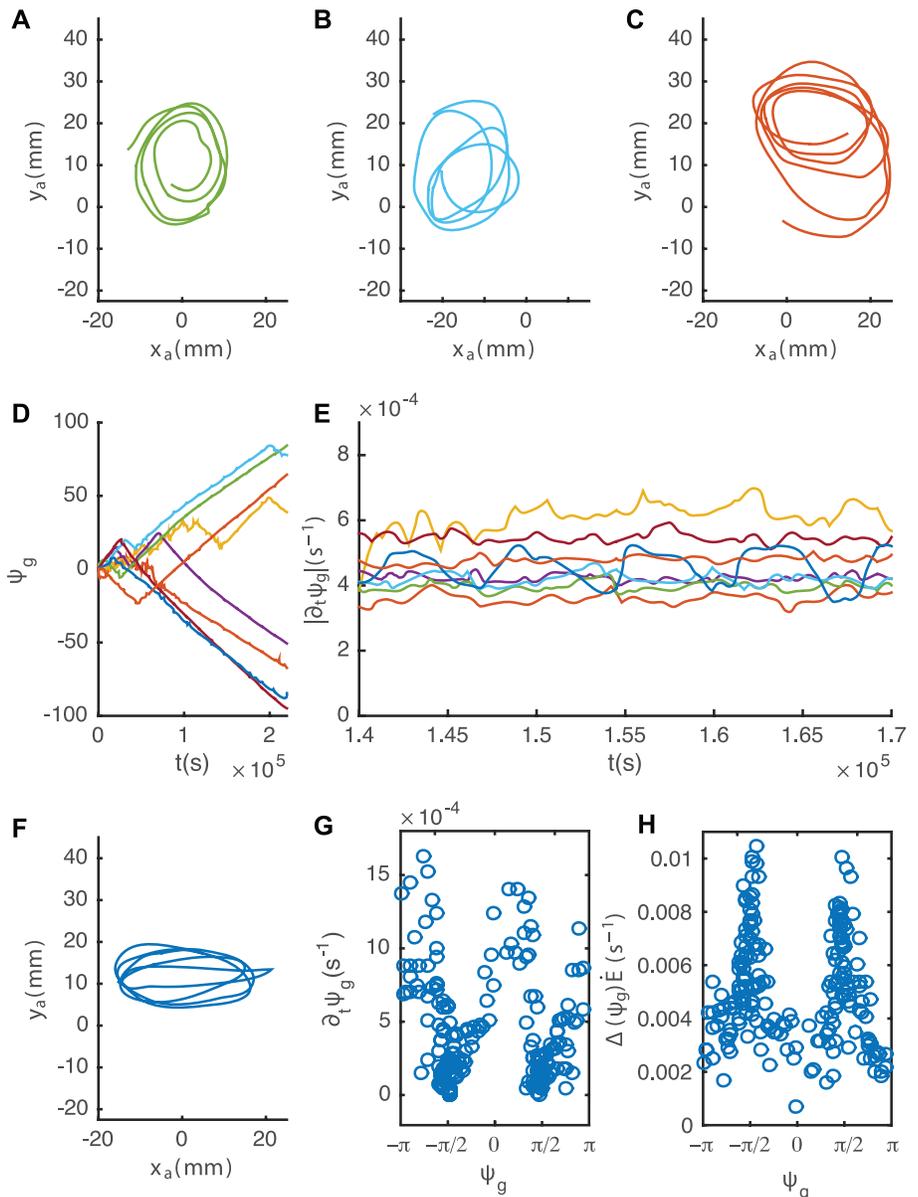}
    \caption{ {\bf A. B. C. and F. Four nutation patterns produced by different plants in the horizontal plane during 63 hours, taken from Stolarz et al~\cite{Stolarz2014}.} D. The principal direction of differential growth $\psi_g(t)$ as a function of time. E. Evolution of $\frac{d\psi_g(t)}{d t}$ as a function of $t$. F. An elliptical pattern produced by a plant in the horizontal plane, not exhibiting torsion. G. $\frac{d\psi_g(t)}{d t}$ as a function of $\psi_g(t)$, showing dips at around $\psi_g=\pm\pi/2$, predicting an elliptical pattern. H.$\Delta(\psi_g(t),t)\dot{E}(t)$ as a function of $\psi_g(t)$, exhibiting peaks at $\psi_g=\pm\pi/2$, in line with the underlying mechanism for an elliptic pattern. }
    \label{exp}
   \end{center}
\end{figure}

We now analyze an existing dataset of apical movements of 8 {\it Arabidopsis thaliana} plants published by Stolarz et al. \cite{Stolarz2014} (see Figure~\ref{exp}A, B, C and F for examples of measured apical patterns.
Most of the observed patterns are elliptical. 
We apply equations~\ref{psig45} and~\ref{dpsig_de} on all 8 measurements of the apical tip in the horizontal plane $(x_a,y_a)$,  extracting $\psi_g(t)$ (shown in Figure~\ref{exp}D)  and $\Delta(\psi_g(t),t)\dot{E}(t)$. 
After $\sim 20$ hours most plants exhibit a linear behavior, equivalent to a constant time derivative $\frac{d\psi_g(t)}{d t}$. Averaging over time and over the different plants results in $\langle \frac{d\psi_g(t)}{d t} \rangle = 4.6\times 10^{-4}\pm 7 \times 10^{-5}s^{-1}$, and substituting this in equation~\ref{tr} gives the time taken for a full rotation of the differential growth direction, $T_r=230\pm40$ min.
At this point we focus on a single curve (plotted in dark blue), where no torsion has been observed. Examining $\partial_t \psi_g$ for closely (plotted in Fig.~\ref{exp}E) we identify oscillations. As mentioned earlier, the model predicts the maxima to be $T_r/2$ apart, representing opposite points along the apical curve. Indeed we find that the average time between every other maximum is $T_o=215\pm25$ min, in agreement with $T_r$ found from the average value of $\partial_t \psi_g$. Moreover, since this plant does not exhibit torsion, one can plot $\frac{d\psi_g(t)}{d t}$ and $\Delta(\psi_g(t),t)\dot{E}(t)$ as a function of $\psi_g(t)$ (shown in Fig.~\ref{exp}G and H). As predicted from the model, we find the minima of $\frac{d\psi_g(t)}{d t}$ and the maxima of  $\Delta(\psi_g(t),t)\dot{E}(t)$ situated at $\psi_g=\pm\pi/2$, representing the farthest sides of the ellipse (on the right and left).

\subsection*{Discussion}

A clear analysis has now been  carried out of the kinematics of differential growth outside of the plane of curvature, and  its implications on plant movement. This is an important step because it shows how a  classical measurement, here the position of the tip in the horizontal plane, is insufficient to provide a clear picture of the relation between observed movements and the underlying growth mechanisms.

Furthermore this study shows that the kinematics of the out of plane curvature can be described  as a simple extension  of the kinematics relating curvature in the plane and differential growth \cite{Bastien2014}. By projecting the differential growth on the planes parallel and perpendicular to the plane of curvature, only one supplementary equation is necessary to describe the full kinematics. This equation relates the orientation of the curvature and the growth in the perpendicular plane. The amplitude of the curvature is modified  by the difference in growth rate between the two sides of the organ in the plane of curvature. The orientation of curvature in space is then modified by the differential growth in the perpendicular plane. Only three parameters are necessary to account for the full movement: i. the elongation rate along the median line $\dot{E}(s,t)$ sets the time scale of the movement, ii. the distribution of the  differential growth in this direction $\Delta(\psi_g(s,t))(s,t)$ and iii. the principal direction of differential growth $\psi_g(s,t)$, the direction in which the differential growth is maximal and . The description of this geometrical framework has been neglected to date, and is a central step to unravel the  relation between differential growth and  nutation.

In both the case of gravitropism~\cite{Bastien2014} and nutation, the destabilizing effects of growth on the movement are regulated by proprioception. The autonomous capacity of plants to control and regulate their own shape is reinforced as a central element of postural control. 
During gravitropic movements, it has been shown experimentally that effects due to growth could be neglected due to the strong influence of proprioception~\cite{Bastien2014}. It is then expected that this regulation is sufficient to avoid the effects due to growth during nutation.

The position of the apical tip in the the horizontal plane, perpendicular to gravity, has been central to the study of kinematics. The relevance of this measure has never been clearly discussed and the underlying hypotheses have remained hidden. In particular the relation between the movement of the apical tip and  the dynamics of the differential growth, the motor of the movement, is difficult to extract because the full shape of the plants remains unknown. A simple set of hypotheses needs to be properly stated to constrain the relation between shapes and movements. The whole organ is now considered as a block that undergoes the same variation all along the organ. Despite the simplification of the problem, this has proved useful to unravel the underlying dynamics of the differential growth, retaining the general observed behavior.  As  plants tend to align their curvature orientation $\psi_c$ with the principal direction of growth $\psi_g$,  the pattern observed in the horizontal plane can remain a marker of growth. Common observed patterns, like the circle or the ellipse, are then directly related to different oscillating patterns of differential growth. Furthermore, simple input such as an oscillation of the principal direction of growth, can produce robust, stable stereotypical patterns independently of the initial conditions. 

Minimal regulation of the movement is necessary to achieve commonly observed patterns like the circle or the ellipse. If measurements in the horizontal plane are useful to understand the kinematics of nutation, they are limited in their scope of analysis. Future studies allowing the measurements of proper 3D kinematics should provide a better understanding of the dynamics of differential growth, and give the exact validity of the measurements performed in the horizontal plane.

\section*{Supporting Information}

The following dimensionless variables are used:
\begin{equation}
\begin{cases}
\dot{E}(s,t)t\rightarrow t'\\
C(s,t)R\rightarrow C'(s,t),
  \end{cases}
\end{equation}
In order to simplify the writing of the equation, the prime indices are dropped. A more compact form to write derivative is also used
\begin{equation}
\begin{cases}
\frac{\textstyle d x}{\textstyle d y}\rightarrow d_y x\\
\frac{\textstyle \partial x}{\textstyle \partial  y}\rightarrow \partial  _y x\\
\frac{\textstyle D x}{\textstyle D  y}\rightarrow D_y x\\
  \end{cases}
\end{equation}

\subsection{Analysis of the movements induced by the principal direction of growth}

Substituting equations~15 and~16 in equations~13 and~14, yields the set of equations:
\begin{equation}\label{dtc_cg}
\partial_{t} C(s,t)=\cos(\psi_g(s,t)-\psi_c(s,t)),
\end{equation}
\begin{equation}\label{dtpsic_cg}
\partial_{t} \psi_c(s,t)=\frac{\sin(\psi_g(s,t)-\psi_c(s,t))}{C(s,t)}.
\end{equation}
Following H1, the dynamics does not depend on the local position along the organ, $C(s,t)\rightarrow C(t)$ and $\psi_c(s,t)\rightarrow\psi_c(t)$ . In following, in order to simplify the writing of the equation, the dependence in time is dropped.
The following set of initial conditions is considered:
\begin{equation}\label{ic_psic}
\psi_c(0)=\psi_c^0
\end{equation}
\begin{equation}\label{ic_c}
C(0)=C_0.
\end{equation}
This accounts for an organ curved in a single plane.
The following change of variables is considered
\begin{equation}
u=\psi_c-\psi_g,
\end{equation}
and substituting in equations~\ref{dtc_cg} and \ref{dtpsic_cg} yields:
\begin{equation}\label{dc_u}
\partial_t C=\cos(u)
\end{equation}
\begin{equation}\label{du}
\partial_t u - \partial_t\psi_g = -\frac{\sin(u)}{C}.
\end{equation}
Rearranging equation~\ref{du} gives and expression for C:
\begin{equation}
C=-\frac{\sin(u)}{\partial_t u-\partial_t\psi_g},
\end{equation}
which we substitute in equation~\ref{dc_u}, yielding 
\begin{equation}
-\frac{\partial_tu}{\partial_tu-\partial \psi_g}\cos u+\frac{\partial_t^2u-\partial_t^2\psi_g}{\left(\partial_t u-\partial_t\psi_g\right)^2}\sin u=\cos u, 
\end{equation}
or alternatively after rearranging:
\begin{equation}
\left(-2\partial_t u+\partial_t \psi_g\right) \cot u+\frac{\partial_t^2u-\partial_t^2\psi_g}{\partial_t u-\partial_t\psi_g}= 0\label{upsi}.
\end{equation}

\subsubsection{Constant principal direction of growth, $\psi_g=0$\label{Constant}}
In the simplest case, where the principal direction of growth is constant,  and arbitrarily chosen to take the value 0:
\begin{equation}
\psi_g=0,
\end{equation}
resulting in $u=\psi_c$, and equation~\ref{upsi} can be rewritten as
\begin{equation}
-2\partial_t \psi_c \cot \psi_c+\frac{\partial_t^2\psi_c}{\partial_t \psi_c}=0.
\end{equation}
Integrating over time yields:
\begin{equation}
-2\log\left(\sin \psi_c \right)+\log\left(\partial_t \psi_c\right)+\tilde{K_1}=0,
\end{equation}
where $\tilde{K_1}$ is an integration constant. Taking an exponential yields:
\begin{equation}\label{dpsic_k1}
\partial_t \psi_c=K_1\sin^2\psi_c,
\end{equation}
where $K_1 = e^{-\tilde{K_1}}$. An additional integration over time results in:
\begin{equation}
\label{psiK}
\cot \psi_c=K_1t+K_2, 
\end{equation}
where $K_2$ is another integration constant. Rearranging results in:
\begin{equation}\label{psic_k1k2}
\psi_c=\arctan\left(\frac{1}{K_1t+K_2}\right).
\end{equation}
Substituting the initial condition in equation \ref{ic_psic}
\begin{equation}
\psi_c(t=0) = \psi_c^0=\arctan\left(\frac{1}{K_2}\right), 
\end{equation}
yields the value of $K_2$:
\begin{equation}\label{K2}
K_2=\cot\psi_c^0.
\end{equation}
We proceed to extract $K_1$, substituting equation~\ref{dpsic_k1} in equation~\ref{dtpsic_cg}:
\begin{equation}
K_1\sin^2\psi_c=-\frac{\sin\psi_c}{C}.
\end{equation}
Substituting the initial conditions stated in equations~\ref{ic_psic} and~\ref{ic_c} yields:
\begin{equation}
K_1\sin^2\psi_c^0=-\frac{\sin\psi_c^0}{C_0},
\end{equation}
\begin{equation}\label{K1}
K_1=-\frac{1}{C_0\sin\psi_c^0}.
\end{equation}
We now substitute the expressions for $K_1$ and $K_2$ (found in equations~\ref{K1} and~\ref{K2}) in equation~\ref{psic_k1k2}, leading to the final expression for $\psi_c(t)$:
\begin{equation}
\psi_c=\arctan\left(\frac{1}{\cot\psi_c^0-t/(C_0\sin\psi_c^0)}\right).
\end{equation}
 Note that for long times, $t\rightarrow \infty$, the direction of maximal curvature is equal to the principal direction of growth, $\psi_c = \psi_g = 0$. 

\subsubsection{ Temporal linear variation of the principal direction of growth, $\psi_g=\omega t$ \label{linear}}
We continue with the linear case
\begin{equation}
\psi_g=\omega t.
\end{equation}
Equation~\ref{upsi} now takes the form:
\begin{equation}
\left(-2\partial_t u+\omega\right) \cot u+\frac{\partial_t^2u}{\partial_t u-\omega}=0.
\end{equation}
Integrating over time yields (with $K'$ is an integration constant):
\begin{equation}
-2\log \sin u+\log\left(\partial_tu-\omega\right)=K'-\omega \int^t \cot u dt',
\end{equation}
and raising to an exponent and rearranging yields:
\begin{equation}\label{du_omega}
\partial_tu=K e^{\textstyle -\omega \int^t \cot u dt'}\sin^2 u+\omega, 
\end{equation}
where $K=e^{K'}$. We substitute equation~\ref{du_omega} in equation~\ref{dtpsic_cg}, and use the initial conditions in equations~\ref{ic_psic} and~\ref{ic_c}, leading to:
\begin{equation}
K\sin^2\psi_c^0=-\frac{\sin\psi_c^0}{C_0},
\end{equation}
and rearranging this yields the integration constant $K$:
\begin{equation}
K=-\frac{1}{C_0\sin\psi_c^0}.
\end{equation}
In order to solve the non linear equation~\ref{du_omega}, the following change of variable is considered:
\begin{equation}
v=\cot u,
\end{equation}
with
\begin{equation}
dv=-\frac{du}{sin^2u}.
\end{equation}
With this change of variables equation~\ref{du_omega} now takes the form:
\begin{equation}
\partial_tv=-K_1\  e^{\textstyle -\omega \int^t v dt'}-\omega\left(1+v^2\right), 
\end{equation}
and taking the logarithm yields:
\begin{equation}
\log\left(\partial_t v+\omega\left(1+v^2\right)\right)=\omega \int^t v dt'+K_1.
\end{equation}
Taking the derivative of time yields:
\begin{equation}
\frac{\partial_t^2v+2\omega v\partial_tv}{\partial_tv+\omega\left(1+v^2\right)}=\omega v,
\end{equation}
and after rearranging:
\begin{equation}
\partial_t^2v+\omega v\partial_tv-\omega^2 v(1+v^2)=0.
\end{equation}
We now introduce a last change of variables:
\begin{equation}
\tau=\omega t,
\end{equation}
leading to the following form:
\begin{equation}\label{neqS}
\partial_{\tau}^2v+ v\partial_{\tau} v- v(1+v^2)=0.
\end{equation}
This equation is then the solution, where we remind that $v = \cot{\left( \psi_c - \psi_g\right)}$.
The phase field of this equation is given for different values of the initial condition. Stable orbits are observed in Figure~S1.
The movement of a single element displays a stable orbit, the  periodicity of which is given by the direction of the differential growth $\omega$. Furthermore the stability and periodicity are independent of the initial conditions. This means that even when the rotation is  not centered around the base of the organ, the pattern remains stable and the periodicity is still given by the internal oscillator.

\begin{figure}[htbp]
  \begin{center}
    \includegraphics[width=0.7\textwidth]{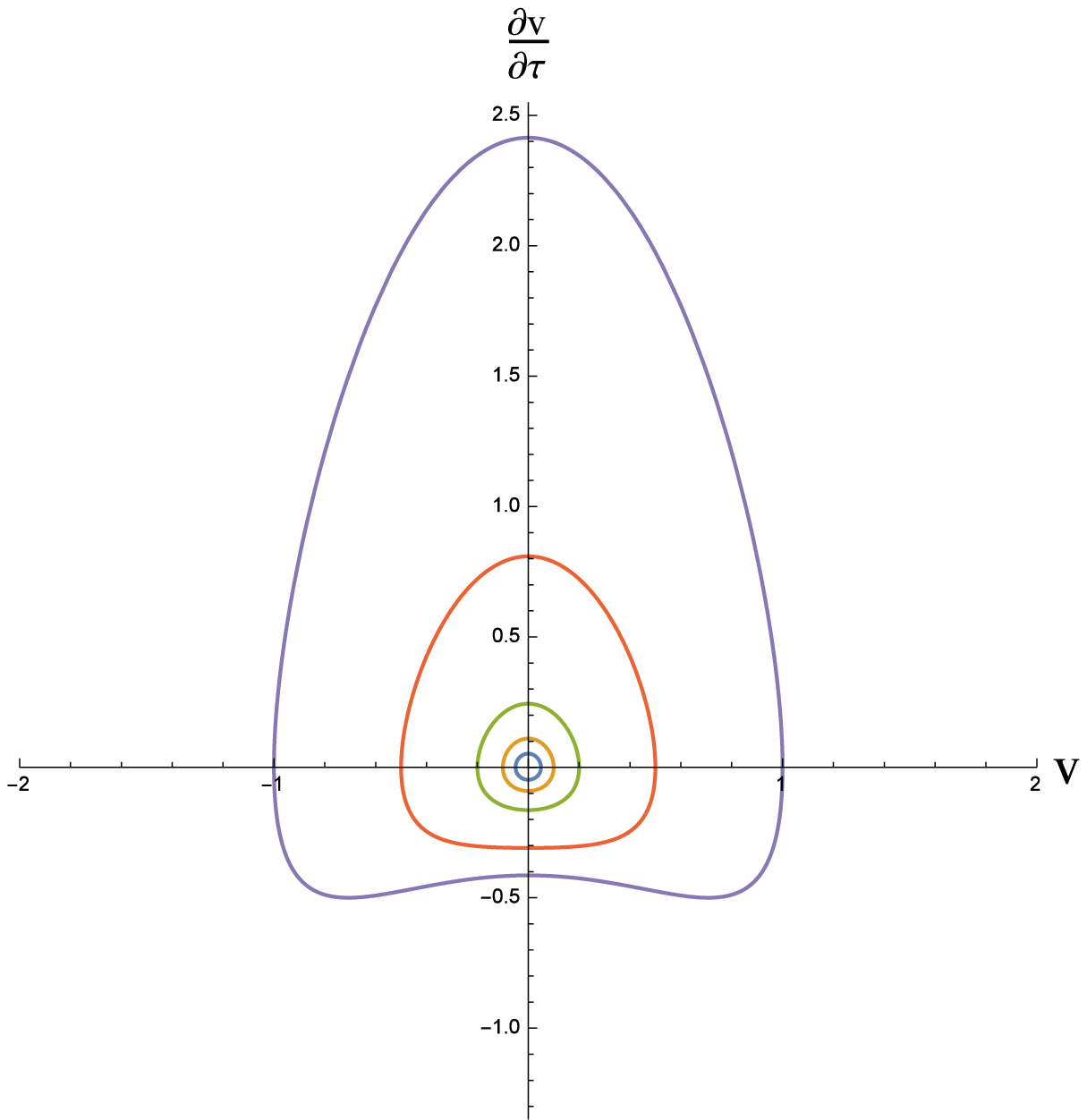}
\captionsetup{labelformat=empty}
    \caption{{\bf Figure S1.}The solutions of equation~\ref{neqS} are represented in the space $(v,\partial_{\tau} v)$, for the following initial conditions blue $v=0.05$, yellow $v=0.1$, green $v=0.2$, red $v=0.5$, dark blue $v=1$ . Orbits are observed and are all travelled during a time $\tau=2\pi$. This means that after a time $\tau=2\pi$, the system comes back to the same initial condition. The orbits are then stable are the same pattern will be displayed at longer times. As $v$ becomes bigger, {it i.e.} the phase between $\psi_g$ and $\psi_c$ is different from $\pi/2$ the dynamics is not symmetrical anymore. However the pattern observed in the apical plane remain the same ({\it e.g.} compared MovieS2 and~S3)}
    \label{exp2}
   \end{center}
\end{figure}

\subsection{Measuring the curvature $C$ from the base-apex distance in the apical plane $\rho_a$ \label{cap}}

\begin{figure}[htbp]
  \begin{center}
    \includegraphics{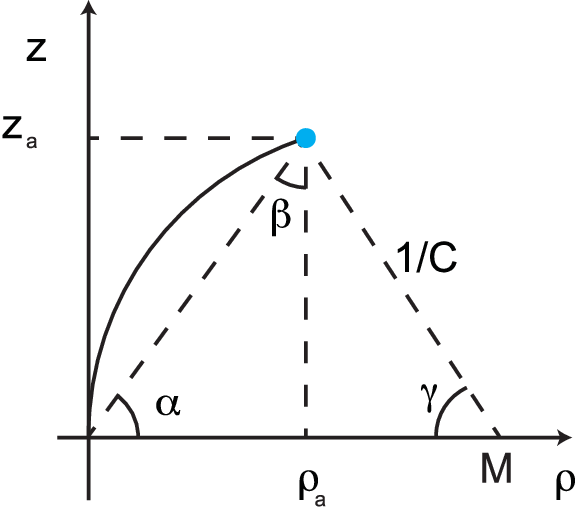}
    
\captionsetup{labelformat=empty}
    \caption{{\bf Figure S2.}Definition of the variables used to measure the curvature in the apical plane. The organ is considered to have a constant curvature $C$ so that the point $M$ on the $rho$ abscissa is the center of the circle described by the organ. The distance from $M$ to the apical tip, $(\rho_a,z_a)$ in the plane $(\rho,z)$, or to the basis of the organ,(0,0), is the radius of curvature $1/C$. }
    \label{geo4}
   \end{center}
\end{figure}

Existing experimental measurements are usually restricted to the apical plane, therefore disregarding important information concerning the conformation of the whole shoot, needed for the analysis presented here. In what follows we show that one can recover the curvature of the shoot $C$ (assumed here to be constant along the shoot) from $\rho_a$, the distance of the apex point from the base in the apical plane, as depicted in Fig.3.C. This can be done through a purely geometric argument, whose details appear in Figure~S2. 

Assuming a constant curvature $C$ along the shoot, we can consider the shoot an arc, part of a circle. The radius of this circle is the inverse of the curvature by definition $R=1/C$, and the length of the arc is just $L$ the length of the shoot. We can therefore calculate the angle defined by the arc, $\gamma$.
\begin{equation}\label{glc}
\gamma = LC.
\end{equation}
Drawing a straight line between the base and apex, we now have an isosceles triangle with equal base angles $\alpha$ and the third angle positioned at the center of the circle of size $\gamma$. Therefore the sum of angles yields:
\begin{equation}\label{triangle_2a}
2\alpha + \gamma = \pi/2
\end{equation}
 The angle between this base-apex line and the projection of the apex to the $\rho_a$ axis is denoted $beta$, defining a right angle triangle, i.e.:
\begin{equation}\label{triangle_ab}
\alpha+\beta = \pi/2.
\end{equation}
Substituting equation~\ref{triangle_ab} in equation~\ref{triangle_2a} results in;
\begin{equation}\label{g2b}
\gamma = 2\beta.
\end{equation}
By definition,
\begin{equation}
\tan\beta = \rho_a/z_a,
\end{equation}
and substituting equations~\ref{g2b} and~\ref{glc} yields:
\begin{equation}
\tan \frac{CL}{2} = \frac{\rho_a}{z_a}.
\end{equation}
For small curvatures $CL\ll 1$, we can approximate $\tan (CL/2) \sim CL/2$, and $z_a \sim L$, leading to
\begin{equation}\label{curvpl}
C = 2L^{-2}\rho_a, 
\end{equation}


\subsection{Measuring $\psi_c$ in the apical plane and its relation to $\psi_g$\label{psap}}
The orientation of the principal direction of curvature of an organ is defined in the apical plane
\begin{equation}\label{psic_atxy}
\psi_c=\arctan\left(\frac{x_a}{y_a}\right).
\end{equation}
The differential element $d\psi_c$ is then given by 
\begin{equation}
d\psi_c=\frac{d(x_a/y_a)}{1+x_a^2/y_a^2},
\end{equation}
or after rearranging:
\begin{equation}\label{dpsic_xy}
d\psi_c=\frac{dy_a}{y_a}\frac{dx_a/dy_a-x_a/y_a}{1+x_a^2/y^2}.
\end{equation}
The distance from the apex to the base in the apical plane, $\rho_a$,  is defined by the position of the apex in that plane, $(x_a, y_a)$, and substituting equation~\ref{curvpl} results in:
\begin{equation}\label{a2c2}
\frac{L^{4}}{4}C^2 = \rho_a^2 = x_a^2+y_a^2. 
\end{equation}
The differential is then:
\begin{equation}\label{a3c3}
\frac{L^{4}}{4} CdC=x_adx_a+y_ady_a.
\end{equation}
Dividing this by equation~\ref{a2c2} and rearranging yields: 
\begin{equation}\label{dcc}
\frac{dC}{C}=\frac{dy_a}{y_a}\frac{\left(dx_a/dy_a\right) \left( x_a/y_a\right)+1}{1+x_a^2/y_a^2},
\end{equation}
where we note that the prefactor $\frac{L^{4}}{4}$ has canceled out. Dividing equation~\ref{dtpsic_cg} by equation~\ref{dtc_cg} results in:
\begin{equation}
\psi_g - \psi_c=\arctan\left(\frac{C\partial_t\psi_c}{\partial_tC}\right),
\end{equation}
and substituting equations~\ref{dpsic_xy} and~\ref{dcc} yields:
\begin{equation}
\psi_g - \psi_c=\arctan\left(\frac{dx_a/dy_a - x_a/y_a}{\left(dx_a/dy_a\right)  \left( x_a/y_a \right)+1}\right).
\end{equation}
We now use the identity $\arctan\left( \frac{u-v}{uv +1}\right) = \arctan\left(u\right) - \arctan\left(v\right)$, and substituting equation~\ref{psic_atxy}, yields:
\begin{equation}
\psi_g=\arctan\left(\frac{dx_a}{dy_a}\right).
\end{equation}
This can be expressed in terms of the temporal derivative of $x_a$ and $y_a$,  finally yielding equation~14 in the main text:
\begin{equation}
\psi_g(t)=\arctan\left(\frac{d_t x_a(t)}{d_t y_a(t)}\right).
\end{equation}
\begin{equation}
\psi_g(t)=\arctan\left(\frac{d x_a(t)}{d y_a(t)}\right).
\end{equation}

\subsection{Measuring $\Delta(\psi_g)\dot{E}$ in the apical plane \label{psap3}}
Substituting equations~\ref{dpsic_xy},~\ref{a2c2} and~\ref{a3c3} into equation~12, we obtain the following relation
\begin{equation}
\Delta(\psi_g(s,t))\dot{E}(s,t)=\sqrt{4L^{-4}\frac{(x_a\partial_tx_a+y_a\partial_ty_a)^2}{x_a^2+y_a^2}R^2+4L^{-4}(x_a^2+y_a^2)\left(\frac{y_a\partial_tx_a - x_a\partial_ty_a}{y_a^2+x_a^2}\right)^2R^2},
\end{equation}
which yields
\begin{equation}
\Delta(\psi_g(s,t))\dot{E}(s,t)=2L^{-2}R\sqrt{\frac{(x_a\partial_tx_a+y_a\partial_ty_a)^2+\left(y_a\partial_tx_a - x_a\partial_ty_a\right)^2}{x_a^2+y_a^2}}.
\end{equation}
and after rearranging
\begin{equation}
\Delta(\psi_g(s,t))\dot{E}(s,t)=2L^{-2}R\sqrt{\partial_tx_a(t)^2+ \partial_ty_a(t)^2}.
\end{equation}

\bibliography{nutation}
\end{document}